\documentstyle[preprint,aps,prb]{revtex}
\begin{document}
\tightenlines
\draft
\title{Final State Effects in the high $q$ response 
of $^3$He--$^4$He mixtures}
\author{F.Mazzanti$^{1,2}$}
\address{$^1$ Institut f\"ur Theoretische Physik, 
        Johannes Kepler Universit\"at Linz, \\
        A-4040 Linz, Austria. \\
        e--mail:\verb+ mazzanti@chioma.tphys.uni-linz.ac.at+}
\address{$^2$ Departament d'Electr\`onica, Enginyeria La Salle,\\
        Pg.Bonanova 8, Universitat Ramon Llull,\\
        E-08022 Barcelona, Spain. \\
        e--mail:\verb+ mazzanti@salleurl.edu+}

\maketitle

\begin{abstract}
A modified Gersch--Rodriguez formalism describing the leading Final
State Effects in the high momentum transfer response of low
concentration $^3$He--$^4$He mixtures is presented and discussed. The
leading corrections to the Impulse Approximation are expressed in
terms of the interatomic potentials and the semidiagonal two--body
density matrices of both the mixture and its boson--boson
approximation, in which the $^3$He atoms are replaced by bosons of the
same mass and at the same partial density. Numerical calculations of
the Final State Effects functions of $^3$He and $^4$He are finally
presented and discussed.  
\end{abstract}

\pagebreak


The dynamic structure function $S(q,\omega)$ is known to contain the
maximum information a neutron scattering experiment can provide about
the structure of a quantum fluid. Many
theoretical~\cite{GerRodSmi,Rinat-1,Glyde-1,MazzBoPo-1} and
experimental~\cite{CowWod,MarSweWod,SosSnoSok-1,KHAnd,RTAzuah} work
designed to study the Bose--Einstein condensate in superfluid $^4$He
has been done in the last thirty years, and much information regarding
its quantum properties have been gathered so far. An alternative
system in which Bose--Einstein condensation sets in is the isotopic
$^3$He--$^4$He mixture, stable at low temperatures and concentrations,
and presenting much richer static and dynamic properties due to the
strong interaction between bosons ($^4$He) and fermions
($^3$He). However, $^3$He atoms are high neutron absorbers and the
experimental measurement of the high $q$ dynamic structure function of
systems containing $^3$He is technically difficult. For this reason,
only one measurement of the high $q$ response of the mixture has been
reported up to the date~\cite{WanSok}, and the analysis of the data
led to surprising conclusions such as a condensate fraction value
$n_0\approx 0.2$, in clear disagreement with almost all theoretical
predictions which are usually close to $n_0\approx
0.1$~\cite{BorPoFab,BonCep}.  However, and despite the existent
discrepancies between theory and experiment, no new measurements of
the high momentum transfer response of the mixture have been performed
yet, and hence the question of what this value really is still remains
open. Consequently, theoretical investigation of the dynamic structure
function of the mixture and further analysis of the measured data is
well justified.

The theoretical analysis of the $q\to\infty$ response of the mixture
is however slightly more involved that in the pure $^4$He case due to
the presence of a fermionic component. In fact, most of the formalisms
that have been used in the past to describe Final State Effects (FSE)
in pure $^4$He rely, in one way or another, on the presence of
long--range order in the the one--body density matrix $\rho_1(r)$.
Long range order makes $\rho_1(r)$ reach a finite $r\to\infty$ value
without changing sign, and thus allowing a cumulant expansion of the
Fourier transform of $S(q,\omega)$ around its high $q$ limit
$\rho_1(tq/m)$. This is not the case when dealing with fermionic
components, and therefore none of these formalisms can be directly
applied to analyze the high $q$ response of $^3$He without further
modifications.  In this work we address this problem and present a new
formalism designed to describe FSE in the response of the
$^3$He--$^4$He mixture, where a $^3$He component is present but
dilute.

In the mixture and when the transferred momentum is high enough for
the incoherent approximation~\cite{Sokmomdist,Glyde-94} to hold, the
dynamic structure function becomes
\begin{equation}
S(q,\omega) = \sigma_4 \left(1-x_3\right) S^{(4)}(q,\omega) + 
\sigma_3 x_3 S^{(3)}(q,\omega) \ ,
\label{intro-b2}
\end{equation}
where $x_3$ is the $^3$He concentration, and $\sigma_3$
and $\sigma_4$ are the cross sections of the separate scattering
processes ($\sigma_3=5.61$ and $\sigma_4=1.34$ in units of
barns~\cite{Sears-86}). Each of the individual responses are the
Fourier transform of their corresponding density--density correlation
factor
\begin{equation}
S^{(\alpha)}(q,t) = \frac{1}{N_\alpha} \sum_{j=1}^{N_\alpha}
\left\langle e^{-i{\bf q}\cdot{\bf r}_j} 
e^{iHt} e^{i{\bf q}\cdot{\bf r}_j} e^{-iHt} \right\rangle 
\label{intro-b3}
\end{equation}
where $\alpha=3$ for $^3$He and $\alpha=4$ for $^4$He. In this
expression, $H$ stands form the Hamiltonian 
\begin{equation}
H = -{1\over 2m_4} \sum_{j\in 4} \nabla_j^2 
-{1\over 2m_3} \sum_{j\in 3} \nabla_j^2 + {1\over 2}
\sum_{\alpha,\beta=3,4} \sum_{i\in\alpha \atop j\in\beta} 
V^{(\alpha,\beta)}(r_{ij})
\label{intro-b4}
\end{equation}
corresponding to a mixture in which a $^3$He and a $^4$He components
coexists and interact through pairwise local and central potentials
$V^{(\alpha,\beta)}(r)$. 

Actually, $S^{(\alpha)}(q,t)$ is the starting point from where a high
$q$ expansion of the response is performed, and this is done acting
with the two translation operators $\exp(-i{\bf q}\cdot{\bf r}_j)$ and
$\exp(i{\bf q}\cdot{\bf r}_j)$ on the time evolution operator
$\exp(-iHt)$ appearing in between. This leads to
\begin{equation}
S^{(\alpha)}(q,t)\!=\! \frac{1}{N_\alpha} e^{i\omega_q^{(\alpha)} t}
\sum_{j=1}^{N_\alpha} \left\langle e^{it(H+L_j^{(\alpha)})}
e^{-itH} \right\rangle \ ,
\label{series-b1}
\end{equation}
where 
$L_j^{(\alpha)}\!=\!{\bf v}^{(\alpha)}\!\cdot\!{\bf p}_j$, 
is the projection of the momentum of
particle $j$ along the direction of the recoiling velocity 
${\bf v}^{(\alpha)}={\bf q}/m_\alpha$ and
$\omega_q^{(\alpha)}\!=\!q^2/2m_\alpha$ may be
understood as its kinetic energy when the momentum transferred to the
system is high. Taking into account the symmetries of the ground state
wavefunction, introducing an identity in the form ${\bf
1}\!=\!\exp(-iL_j^{(\alpha)}t) \exp(iL_j^{(\alpha)}t)$ and moving the
time evolution operator to the left, the density--density correlation
factor becomes
\begin{equation}
S^{(\alpha)}(q,t)\!=\!e^{i\omega_q^{(\alpha)} t} \left\langle 
{\mathcal{C}}^{(\alpha)}(t) e^{itL_1^{(\alpha)}} \right\rangle \ ,
\label{series-b2}
\end{equation}
where ${\mathcal{C}}^{(\alpha)}(s)$ is the operator carrying the
effects of the FSE, and reads
\begin{equation}
{\mathcal{C}}^{(\alpha)}(t) \!\equiv\! e^{-itH} e^{it(H+L_1^{(\alpha)})} 
e^{-itL_1} \ .
\label{series-b3}
\end{equation}

In the high $q$ limit, $v^{(\alpha)}$ is large and $t$ is short while
their product remains of order unity. Therefore, one is allowed to
write the response as a function of $s=v^{(\alpha)}t$ instead of $t$,
and expand it in inverse powers of the recoiling velocity
$v^{(\alpha)}$. Defining a new Hamiltonian
${\mathcal{H}}=H/v^{(\alpha)}$, Eqs.~(\ref{series-b2})
and~(\ref{series-b3}) become
\begin{mathletters}
\begin{eqnarray}
S^{(\alpha)}(q,s) & = & 
e^{is\omega_q^{(\alpha)}\!/v^{(\alpha)}}
\left\langle {\mathcal{C}}^{(\alpha)}(s) 
e^{is\hat{\bf v}^{(\alpha)}\cdot{\bf p}_1} \right\rangle 
\label{series-b4a} \\
{\mathcal{C}}^{(\alpha)}(s) & = & 
e^{-is{\mathcal{H}}} 
e^{is({\mathcal{H}} + \hat{\bf v}^{(\alpha)}\cdot{\bf p}_1)}
e^{-is{\bf v}^{(\alpha)}\cdot{\bf p}_1} \ ,
\label{series-b4b}
\end{eqnarray}
\end{mathletters}
where $\hat{\bf v}^{(\alpha)}$ is the unit vector pointing in the
direction of ${\bf v}^{(\alpha)}$. Direct differentiation
of~(\ref{series-b4a}) leads to the following first order differential
equation satisfied by ${\mathcal{C}}^{(\alpha)}(s)$
\begin{equation}
\frac{d}{ds}{\mathcal{C}}^{(\alpha)}(s) 
\equiv \dot{\mathcal{C}}^{(\alpha)}(s) = 
i\left[ \hat{\mathcal{H}}(s) - {\mathcal{H}} \right]
{\mathcal{C}}^{(\alpha)}(s) 
\label{series-b5} 
\end{equation}
which must be solved together with the initial condition
${\mathcal{C}}^{(\alpha)}(0)\!=\!{\bf 1}$, the latter being implicit
from the definition of ${\mathcal C}^{(\alpha)}(s)$ in
Eq.~(\ref{series-b4b}). In Eq.~(\ref{series-b5}), $\hat{\mathcal{H}}(s)$
is an effective Hamiltonian that results from the action on
${\mathcal{H}}$ of the different exponential operators appearing in
${\mathcal{C}}^{(\alpha)}(s)$
\begin{equation}
\hat{\mathcal{H}}(s) \equiv 
e^{-is{\mathcal{H}}}
e^{is({\mathcal{H}}+\hat{\bf v}^{(\alpha)}\cdot{\bf p}_1)}
{\mathcal H}
e^{-is({\mathcal{H}}+\hat{\bf v}^{(\alpha)}\cdot{\bf p}_1)}
e^{is{\mathcal{H}}} \ .
\label{series-b6}
\end{equation}

Different representations of the density-density correlation factor
can be obtained from the different forms in which the solution of
Eq.~(\ref{series-b5}) can be organized. One form that has proved 
to be particularly useful in the past  is the cumulant
one~\cite{GerRod,Glyde-94}, which captives the most salient
features of the response and simplifies in the $q\to\infty$ limit
\begin{equation}
{\mathcal{C}}^{(\alpha)}(s) \equiv e^{\epsilon_\alpha \Delta_0(s)}
e^{\epsilon_\alpha^2\Delta_1(s)} e^{\epsilon_\alpha^3\Delta_2(s)} 
\cdots = \prod_{n=1}^\infty \exp\left[\epsilon_\alpha^n 
\Delta_n(s)\right] \ .
\label{series-e3}
\end{equation}
Assuming this analytical form, the first $s$--derivative of
${\mathcal{C}}^{(\alpha)}(s)$ can be easily calculated and plugged
into Eq.~(\ref{series-b5}). Then equating each order in
$\epsilon_\alpha^n$ in both sides of the equation one finds a
hierarchy of coupled differential equations that can be solved for
every $\Delta_n$ once the previous $\Delta_, \Delta_1, \ldots,
\Delta_{n-1}$ are known. The first terms in the chain can be readily
obtained and lead to
$\epsilon_\alpha$
\begin{eqnarray}
{\mathcal{C}}^{(\alpha)}(s) & = & \exp\left[ \,
i\epsilon_\alpha\int_0^s\left(\hat{H}_0(y)-H\right)dy\right]
\exp\left[i\epsilon_\alpha^2\int_0^s\hat{H}_1(y)dy \right]
\nonumber \\ [2mm]
& & \times\exp\left[\epsilon_\alpha^3\left( i\int_0^s\hat{H}_2(s') ds' 
-\int_0^s\int_0^{s'} \left[ \hat{H}_1(s'), \hat{H}_0(s'')-H \right] 
ds' ds'' \right)\right] \cdots \ ,
\label{series-f3}
\end{eqnarray}

Alternatively, ${\mathcal C}(s)$ can be casted as an additive series
setting
\begin{equation}
e^{\epsilon_\alpha^n \Delta_n(s)} \equiv 1 +
\epsilon_\alpha^n \Gamma_n(s)
\label{series-f4}
\end{equation}
where $\Gamma_n(s)$ is to the lowest order independent of
$\epsilon_\alpha$. In this case one finally arrives at the general
solution
\begin{eqnarray}
{\mathcal{C}}^{(\alpha)}(s) & =  &
\left[1+\epsilon_\alpha\Gamma_1(s)\right]
\left[1+\epsilon_\alpha^2\Gamma_2(s)\right]
\left[1+\epsilon_\alpha^3\Gamma_3(s)\right] \cdots
\nonumber \\ [2mm]
& \equiv & e^{i\epsilon_\alpha \int_0^s\left(\hat{H}_0(s')-H\right) ds'} + 
\left( e^{i\epsilon_\alpha^2\int_0^s \hat{H}_1(s') ds'}-1\right) + 
\ldots \ . 
\label{series-f5}
\end{eqnarray}

Equations~(\ref{series-f3}) and~(\ref{series-f5}) are different
expressions of the same solution. Both of them can be used in
Eq.~(\ref{series-b4a}) to generate a representation of the
density--density correlation factor in which each order in
$\epsilon_\alpha$ is separately displayed. As a matter of fact, both
forms coincide in the $\epsilon_\alpha\to 0$ limit, where only the
zero order term contributes. In this case
${\mathcal{C}}^{(\alpha)}(s)=1$ and $S^{(\alpha)}(q,s)$ reduces to the
Impulse Approximation
\begin{equation}
S_0^{(\alpha)}(q,s) = e^{i\omega_q^{(\alpha)}/v^{(\alpha)}} \left\langle
e^{is\hat{\bf v}^{(\alpha)}\cdot{\bf p}_1} \right\rangle \equiv
e^{i\omega_q^{(\alpha)}/v^{(\alpha)}} \frac{1}{\rho_\alpha} \,
\rho_1^{(\alpha)}(s)
\label{series-f6}
\end{equation}
which is known to dominate the $q\to\infty$ behavior of the response.
At high but finite $q$, $\epsilon_\alpha$ is small and thus the
leading corrections to the IA are already captived by the first term
in Eq.~(\ref{series-f5}), and so all other terms may be
discarded. This leads to the following approximation for the response
\begin{equation}
S_1^{(\alpha)}(q,s) = e^{is\omega_q^{(\alpha)}/v^{(\alpha)}}
\left\langle 
e^{i\epsilon_\alpha \int_0^s \left(\hat{H}_0(s')-H\right)ds'}
e^{i{\bf s}\cdot{\bf p}_1} \right\rangle \ .
\label{series-g2}
\end{equation}

The difference $\hat{H}_0(s')-H$ in the argument of the exponential
can be written in terms of the interatomic potentials alone.  In the
general case considered here in which two different species coexist
and interact through pairwise local and central forces, the argument
of the exponential reduces to
\begin{eqnarray}
\hat{H}_0(s) - H & = & 
e^{i{\bf s}\cdot{\bf p}_1} H e^{-i{\bf s}\cdot{\bf p}_1} - H
\nonumber \\
& = & H + \sum_{\beta=3,4} 
\sum_{j\in\beta} \left( e^{i{\bf s}\cdot{\bf p}_1}
V^{(\alpha\beta)}(r_{1j}) e^{-i{\bf s}\cdot{\bf p}_1} 
- V^{(\alpha\beta)}(r_{1j}) \right) - H
\nonumber \\
& = & \sum_{\beta=3,4} \sum_{j\in\beta}\left( 
V^{(\alpha\beta)}({\bf r}_{1j}+{\bf s})
-V^{(\alpha\beta)}(r_{1j}) \right)
\equiv \sum_{\beta=3,4} \sum_{j\in\beta} 
\Delta V^{(\alpha\beta)}({\bf r}_{1j}, {\bf s}) \ ,
\label{series-g3}
\end{eqnarray}
where in the last line use has been made of the definition of the
potential difference
\begin{equation}
\Delta V^{(\alpha\beta)}({\bf r}_{ij}, {\bf r}') \equiv 
V^{(\alpha\beta)}({\bf r}_{ij}+{\bf r}') - 
V^{(\alpha\beta)}(r_{ij}) \ . 
\label{series-g4}
\end{equation}
Moreover, the leftmost exponential operator inside the expectation
value in Eq.~(\ref{series-g2}) is diagonal in configuration space, and
therefore $S_1^{(\alpha)}(q,s)$ reduces to an integral over the
semidiagonal $N$--body density matrix and the interatomic potentials
\begin{eqnarray}
S^{(\alpha)}_1(q,s) = \frac{1}{N_3!N_4!}
&\,\,& e^{i\omega_q^{(\alpha)}/v^{(\alpha)}} 
\int d{\bf r}^N 
\rho_N({\bf r}_1, {\bf r}_2, \ldots, {\bf r}_N; {\bf r}_1+{\bf s}) 
\nonumber \\ [2mm]
& & \times\exp\left[ \frac{i}{v^{(\alpha)}} \sum_{\beta=3,4} \sum_{j\in\beta} 
\int_0^s ds' \Delta V^{(\alpha\beta)}({\bf r}_{1j},{\bf s}') 
\right] \ .
\label{series-g6}
\end{eqnarray}
At zero temperature $\rho_N$ is defined in terms of the ground
state wavefunction $\Psi_0$ 
\begin{equation}
\rho_N({\bf r}_1, {\bf r}_2, \ldots, {\bf r}_N; {\bf r}'_1) 
\equiv N_3!N_4! \Psi_0^*({\bf r}_1, {\bf r}_2, \ldots, {\bf r}_N)
\Psi({\bf r}'_1, {\bf r}_2, \ldots, {\bf r}_N) \ ,
\label{series-g7}
\end{equation}
where the $N_3$ coordinates corresponding to $^3$He particles
represent both position and spin coordinates. 

Eq.~(\ref{series-g6}) is simple compared to the expression of the
exact response, but it is still very difficult to evaluate due to the
$N$--body quantities entering on it. However, it is in the appropriate
form for a Gersch--Rodriguez cumulant expansion~\cite{GerRod}. While
the original Gersch--Rodriguez formalism deals with expectation values
of time--ordered integrals of operators, the approximations made so
far have brought the response to the {\em Static Background
Approximation}. In this limit, the scattering time is assumed to be so
short and the momentum transfer so high that in essence only the
particle being struck by the incoming neutron moves in the process,
and thus all other particles in the background are assumed to be {\em
frozen} at their positions. Under these circumstances, the series
simplifies and reduces to the general expansion rule
\begin{equation}
\phi({\bf r}_1, {\bf r}'_1) + \int d{\bf r}^N
f({\bf r}_1, {\bf r}_2, \ldots, {\bf r}_N; {\bf r}'_1)
\exp\left[ i\sum_m \int_0^s ds' \theta_m(s') \right]
\equiv W_0 \exp\left[\sum_{n=1}^\infty \omega_n \right] \ ,
\label{series-h1}
\end{equation}
where $\phi({\bf r}_1, {\bf r}'_1)$, $f({\bf r}_1, \ldots, {\bf r}_N;
{\bf r}'_1)$ and $\theta_m(s)$ are arbitrary functions, the latter
also possibly depending on particle coordinates ${\bf r}_1, {\bf
r}'_1, {\bf r}_2, \ldots, {\bf r}_N$. $W_0$ and $\omega_n$ are the
coefficients of the expansion and up to $n=1$ read
\begin{mathletters}
\label{series-h2}
\begin{eqnarray}
W_0 & = & \phi({\bf r}_1, {\bf r}'_1) + 
\int d{\bf r}^N f({\bf r}_1, {\bf r}_2, \ldots, {\bf r}_N; {\bf r}'_1) 
\label{series-h2a} \\ [2mm]
\omega_1 & = & -\frac{1}{W_0} \sum_m \int d{\bf r}^N 
f({\bf r}_1, \ldots, {\bf r}_N; {\bf r}'_1) \left[ 1 - 
\exp\left[i\int_0^s ds'\theta_m({\bf s}') \right] \right] \ ,
\label{series-h2b}
\end{eqnarray}
\end{mathletters}
while expression for higher order terms can be derived following the
steps described in the literature~\cite{GerRod}.  Notice that due to
the logarithmic nature of the expansion, this relation holds only if
$W_0$ has no zeros. In the current case, $W_0$ is related to the
one--body density matrix of the component whose density--density
correlation factor is being analyzed, and whereas $\rho_1^{(4)}(s)$ is
everywhere positive and fulfills the required condition,
$\rho_1^{(3)}(s)$ presents a complex nodal structure that makes the
applicability of the expansion rely on the appropriate choice of
$\phi({\bf r}_1,{\bf r}'_1)$.

When expressions~(\ref{series-h1}),~(\ref{series-h2a})
and~(\ref{series-h2b}) are used to compute the density--density
correlation factor of the $^4$He component of the mixture, one arrives
at the natural extension of the original Gersch--Rodriguez
result. Setting $\phi({\bf r}_1, {\bf r}'_1)=0$, one gets
\begin{equation}
W_0 = \frac{1}{N_3!N_4!} \int d{\bf r}^N 
\rho_N({\bf r}_1, {\bf r}_2, \ldots, {\bf r}_N; {\bf r}_1+{\bf s})
= \frac{1}{\rho_4} \rho_1^{(4)}(s) 
\label{4heresp-b1}
\end{equation}
while $\omega_1$ reads
\begin{eqnarray}
\omega_1 & = & 
-\frac{1}{N_3!N_4!} \frac{\rho_4}{\rho_1^{(4)}(s)}
\sum_{\beta=3,4} \sum_{j\in\beta \atop j\neq 1} \int d{\bf r}^N
\rho_N({\bf r}_1, \ldots, {\bf r}_N; {\bf r}_1+{\bf s})
\left[1-\exp\left(\frac{i}{v^{(4)}} \int_0^s ds'
\Delta V^{(4\beta)}({\bf r}_{1j}, {\bf s}') \right) \right]
\nonumber \\ [2mm]
& = & -\frac{1}{\rho_1^{(4)}} \int d{\bf r}
\rho_2^{(4,4)}({\bf r}, 0; {\bf r}+{\bf s})
\left[1-\exp\left(\frac{i}{v^{(4)}} \int_0^s ds'
\Delta V^{(44)}({\bf r}, {\bf s}') \right) \right]
\nonumber \\ [2mm]
& & -\frac{1}{\rho_1^{(4)}} \int d{\bf r}
\rho_2^{(4,3)}({\bf r}, 0; {\bf r}+{\bf s})
\left[1-\exp\left(\frac{i}{v^{(4)}} \int_0^s ds'
\Delta V^{(43)}({\bf r}, {\bf s}') \right) \right]
\label{4heresp-b2}
\end{eqnarray}
in terms of the $(4,4)$ and $(4,3)$ components of the semi--diagonal
two--body density matrix of the mixture
\begin{equation}
\rho_2^{(\alpha,\beta)}({\bf r}_1, {\bf r}_2; {\bf r}'_1) = 
N_\alpha \left( N_\beta - \delta_{\alpha\beta} \right)
\frac{\int d{\bf r}_3\cdots d{\bf r}_N
\Psi_0^*({\bf r}_1, {\bf r}_2, \ldots, {\bf r}_N)
\Psi_0^({\bf r}'_1, {\bf r}_2, \ldots, {\bf r}_N)}
{\int d{\bf r}^N \left|
\Psi_0^({\bf r}_1, {\bf r}_2, \ldots, {\bf r}_N) \right|^2} \ .
\label{4Heresp-b2b}
\end{equation}

At this level, therefore, S$^{(4)}(q,s)$ is predicted to be the
algebraic product of the IA and the FSE broadening function
\begin{equation}
S^{(4)}_1(q,s) = S^{(4)}_{IA}(q,s) R^{(4)}(q,s) \ ,
\label{4heresp-b3}
\end{equation}
where
\begin{equation}
S^{(4)}_{IA}(q,s) = e^{is\omega_q^{(4)}/v^{(4)}}
\frac{1}{\rho_1^{(4)}} \rho_1^{(4)}(s) 
\label{4heresp-b4a} 
\end{equation}
and 
\begin{eqnarray}
R^{(4)}(q,s) & = & \exp\left[ -\frac{1}{\rho_1^{(4)}(s)}
\int d{\bf r} \rho_2^{(4,4)}({\bf r},0;{\bf r}+{\bf s})
\left[ 1-\exp\left(\frac{i}{v^{(4)}} 
\int_0^s ds'\Delta V^{(44)}({\bf r},{\bf s}') \right)\right]\right.
\nonumber \\ [2mm]
& - & \!\!\left. \frac{1}{\rho_1^{(4)}(s)} 
\int d{\bf r} \rho_2^{(4,3)}({\bf r},0;{\bf r}+{\bf s})
\left[ 1-\exp\left(\frac{i}{v^{(4)}} 
\int_0^s ds'\Delta V^{(43)}({\bf r},{\bf s}') \right)\right]\right] \ .
\label{4heresp-b4b}
\end{eqnarray}

These results are formally equal to the Gersch--Rodriguez ones used to
compute the high $q$ response of pure $^4$He, the only difference
being the presence of $\rho_2^{(4,3)}$ which is the contribution
coming from the interaction of the $^4$He atoms with the $^3$He atoms
in the mixture. As a matter of fact, in the zero $^3$He concentration
limit this last term cancels and expression~(\ref{4heresp-b4b})
coincides exactly with the one reported in ref.~\cite{GerRod}, thus
stressing that the former is the natural extension of the latter to
the mixture where two different isotopes coexist.

Unfortunately, no cumulant expansion of the $^3$He density--density
correlation factor can be performed as described above. This is
because the zero order cumulant, which is proportional to the $^3$He
one--body density matrix, has infinitely many nodes.  This problem can
be bypassed recalling that in the high momentum transfer limit the
most relevant processes in the scattering are those taking place at
short distances, where dynamical correlations dominate over
statistical ones. This means that in the $q\to\infty$ limit, and
disregarding the effects on the Bose--Einstein condensate, FSE in
boson and fermion systems should look like similar, and that therefore
the contribution of the fermion statistics to the FSE can be
introduced as a small correction to the effect produced by the
dynamical correlations, which are entirely taken into account by a
purely bosonic FSE function. But on the other hand, the IA is known to
substantially depend on the statistics obeyed by the system. The $s$
representation of the IA is proportional to the one--body density
matrix, so it seems that this is the only quantity that really
requires a proper treatment of the statistics. One can consider,
therefore, a factorization of the $N$--body density matrix of the
mixture in the following form
\begin{eqnarray}
\rho_N({\bf r}_1, {\bf r}_2, \ldots, {\bf r}_N; {\bf r}'_1) & = &
\rho_1^{(3)}(r_{11'}) \left[ \frac{1}{\rho_1^B(r_{11'})} 
\rho_N^B({\bf r}_1, {\bf r}_2, \ldots, {\bf r}_N; {\bf r}'_1) \right]
\nonumber \\ [2mm]
& + & \left[ 
\rho_N({\bf r}_1, {\bf r}_2, \ldots, {\bf r}_N; {\bf r}'_1) 
-\frac{\rho_1^{(3)}(r_{11'})}{\rho_1^B(r_{11'})} 
\rho^B_N({\bf r}_1, {\bf r}_2, \ldots, {\bf r}_N; {\bf r}'_1) 
\right] \ ,
\label{resp3he-b2}
\end{eqnarray}
where $\rho_1^B$ and $\rho_N^B$ are the one- and the $N$--body density
matrices of a fictitious mixture (henceforth referred to as the {\em
boson--boson mixture}) in which the $^3$He atoms are replaced by
bosons of the same mass and at the same partial density. Notice that
this replacement does not introduce any singularity in $\rho_N$
because $\rho_1^B(r)$ corresponds to a bosonic phase and thus has no
nodes.

With this prescription, the $^3$He response of the mixture becomes the
sum of two terms
\begin{equation}
S^{(3)}_1(q,s) \equiv S_B^{(3)}(q,s) + \Delta S^{(3)}(q,s) \ ,
\label{resp3he-c1}
\end{equation}
where
\begin{eqnarray}
S_B^{(3)}\!\!& (&\!\!q,s) = e^{is\omega_q^{(3)}/v^{(3)}} 
\frac{1}{\rho_3} \rho_1^{(3)}(s) 
\label{resp3he-c2a} \\ [2mm]
& \times & \!\left[ \frac{\rho_3}{N_3!N_4!} \int d{\bf r}^N
\frac{1}{\rho_1^B(s)} \rho_N^B({\bf r}_1, {\bf r}_2, \ldots, 
{\bf r}_N; {\bf r}_1+{\bf s}) \exp\left({\frac{i}{v^{(3)}}\!\!
\sum_{\beta=3,4}\sum_{j\in\beta \atop j\neq 1} \int_0^s ds'
\Delta V^{(3\beta)}({\bf r}_{1j}, {\bf s}')} \right) \right]
\nonumber
\end{eqnarray}
and
\begin{eqnarray}
\Delta S^{(3)}(q,s) & = & e^{is\omega_q^{(3)}/v^{(3)}} 
\frac{1}{N_3!N_4!} \int d{\bf r}^N \biggl[ 
\rho_N({\bf r}_1, {\bf r}_2, \ldots, {\bf r}_N; {\bf r}_1 + {\bf s}) 
\label{resp3he-c2b} \\ [2mm]
& - & \frac{\rho_1^{(3)}(s)}{\rho_1^B(s)} 
\rho_N^B({\bf r}_1, {\bf r}_2, \ldots, {\bf r}_N; {\bf r}_1 + {\bf s}) 
\biggl] \exp\left({\frac{i}{v^{(3)}}\!
\sum_{\beta=3,4} \sum_{j\in\beta \atop j\neq 1}\! 
\int_0^s ds'\Delta V^{(3\beta)}({\bf r}_{1j}, {\bf s}')}\right) \ .
\nonumber
\end{eqnarray}

In this approximation, Eq.~(\ref{resp3he-c2a}) describes that part of
the response that can be written as the product of the IA, which
enters through $\rho_1^{(3)}(s)$, with a bosonic FSE function, that is
given by the term in square brackets. It is important to notice,
however, that the IA is exactly accounted in this term because the
{\em exact} $^3$He one--body density matrix has been previously
factorized; while the FSE broadening function is evaluated in the
boson--boson approximation. In contrast, $\Delta S^{(3)}(q,s)$ carries
the contribution of all the statistical correlations between particles
$2, 3, \ldots, N_3$ of $^3$He that have not been taken into account in
$\rho_1^{(3)}(s)$. This last terms is expected to introduce small
corrections to $S^{(3)}(q,s)$ that only appear when the response is
computed beyond the IA level.

As before, Eqs.~(\ref{resp3he-c2a}) and~(\ref{resp3he-c2b}) are
difficult to handle due to the presence of $\rho_N$ and
$\rho_N^B$. Nevertheless, the FSE function in~(\ref{resp3he-c2a}) can
be worked out just as in the $^4$He case due to the bosonic nature of
the functions entering on it. Hence one finds
\begin{equation}
S^{(3)}_B(q,s) = S_{IA}^{(3)}(q,s)\,R^{(3)}(q,s) \ ,
\label{resp3he-d1}
\end{equation}
where
\begin{equation}
S_{IA}^{(3)}(q,s) = e^{i\omega_q^{(3)}/v^{(3)}} 
\frac{1}{\rho_3} \rho_1^{(3)}(s)
\label{resp3he-d2a} 
\end{equation}
and
\begin{eqnarray}
R^{(3)}(q,s) & = & \exp\left[ -\frac{1}{\rho_1^B(s)} 
\int d{\bf r} \rho_2^{(3,3)B}({\bf r}, 0; {\bf r}+{\bf s}) 
\left[ 1 - \exp\left(\frac{i}{v^{(3)}} \int_0^s ds'
\Delta V^{(33)}({\bf r}, {\bf s}') \right) \right] \right.
\nonumber \\ [2mm]
& & \left. -\frac{1}{\rho_1^B(s)}
\int d{\bf r} \rho_2^{(3,4)B}({\bf r}, 0; {\bf r}+{\bf s}) 
\left[ 1 - \exp\left(\frac{i}{v^{(3)}} \int_0^s ds'
\Delta V^{(34)}({\bf r}, {\bf s}') \right) \right] \right] \ .
\label{resp3he-d2b}
\end{eqnarray}

The $^3$He additive term $\Delta S^{(3)}(q,s)$ can not be handled in
the same way because in this case the zero order cumulant vanishes
\begin{equation}
W_0 = \frac{1}{N_3!N_4!} \int d{\bf r}^N \left[
\rho_N({\bf r}_1, {\bf r}_2, \ldots, {\bf r}_1+{\bf s}) - 
\frac{\rho_1^{(3)}(s)}{\rho_1^B(s)} 
\rho_N^B({\bf r}_1, {\bf r}_2, \ldots, {\bf r}_1+{\bf s})
\right] = 0
\label{resp3he-d3}
\end{equation}
thus violating the condition imposed on $W_0$. 

The problem of finding a useful prescription for 
$\Delta S^{(3)}(q,s)$ can be solved inspecting the structure of the
two--body density matrices that would enter in the lowest order terms
of a cumulant expansion, as they carry the leading contributions to
the FSE in the $q\to\infty$ limit. When Eq.~(\ref{resp3he-b2}) is
integrated over all particle coordinates but ${\bf r}_1$ and ${\bf
r}_2$, an equivalent factorization of $\rho_2^{(3,\alpha)}$ is
found. The difference between the $^3$He semidiagonal two--body
density matrices of the mixture and their boson--boson approximation
\begin{equation}
\Delta\rho_2^{(3,\alpha)}({\bf r}_1, {\bf r}_2; {\bf r}'_1) = 
\rho_2^{(3,\alpha)}({\bf r}_1, {\bf r}_2; {\bf r}'_1) -
\frac{\rho_1^{(3)}(r_{11'})}{\rho_1^B(r_{11'})}
\rho_2^{(3,\alpha)B}({\bf r}_1, {\bf r}_2; {\bf r}'_1) 
\label{resp3he-e1}
\end{equation}
can be analyzed in the framework of the HNC/FHNC equations starting
from a variational model of the ground state wave function. Careful
inspection of the diagrams entering in~(\ref{resp3he-e1}) reveals that
$\Delta\rho_2^{(3,\alpha)}({\bf r}_1, {\bf r}_2; {\bf r}'_1)$ may be
factorized as follows~\cite{RisCla,FerPhD}
\begin{equation}
\Delta\rho_2^{(3,\alpha)}({\bf r}_1, {\bf r}_2; {\bf r}'_1) = 
\rho_\alpha \rho_1^{(3)}(r_{11'}) 
G^{(3,\alpha)}({\bf r}_1, {\bf r}_2; {\bf r}'_1) - 
\rho_\alpha \rho_{1D}(r_{11'}) 
F^{(3,\alpha)}({\bf r}_1, {\bf r}_2; {\bf r}'_1) \ ,
\label{resp3he-e2}
\end{equation}
where $\rho_{1D}(r_{11'})$ is an auxiliary function that adds the
contribution of all those diagrams linking points $1$ and $1'$ that
are not connected to point $2$ and with no statistical lines starting
or ending at points~$1$ and~$1'$. Function $G^{(3,\alpha)}({\bf r}_1,
{\bf r}'_1; {\bf r}_2)$ and $F^{(3,\alpha)}({\bf r}_1, {\bf r}'_1;
{\bf r}_2)$ are the sum of all other diagrams not contributing to
$\rho_1^{(3)}(s)$ that contain dynamical and statistical lines linking
the external points. Indeed, it can be seen from its diagrammatic
structure that $\rho_{1D}(r_{11'})$ shares many common features with
the one--body density matrix of a purely bosonic liquid, as for
instance it is always positive and its large $r$ value approaches a
constant that would be ascribed to some sort of condensate fraction
value, although it has no physical meaning in this case. As a matter
of fact, numerical calculations show that at low $^3$He concentrations
$\rho_{1D}(r_{11'})$ and $\rho_1^B(r_{11'})$ are nearly identical, and
so that both $\rho_{1D}(r_{11'})$ and $\rho_1^B(r_{11'})$ satisfy the
expansion condition and that either of them can be used as the basic
function $\phi({\bf r}_1, {\bf r}'_1)$ upon which the cumulant
expansion is being built. Choosing $\rho_{1D}(r_{11'})$ as the
starting function, $\Delta S^{(3)}(q,s)$ in Eq.~(\ref{resp3he-c2b})
can be brought to a form suitable for expansion purposes by adding and
subtracting the former to the latter
\begin{eqnarray}
\Delta S^{(3)}(q,s) & = & 
e^{is\omega^{(3)}/v^{(3)}} \frac{1}{\rho_3} 
\Biggl[ \rho_{1D}(s) + \frac{\rho_3}{N_3!N_4!} \int d{\bf r}^N
\biggl(\rho_N({\bf r}_1, {\bf r}_2, \ldots, {\bf r}_N;
{\bf r}_1+{\bf s}) 
\nonumber \\ [2mm]
& - & \frac{\rho_1^{(3)}(s)}{\rho_{1D}(s)}
\rho_N^{(3,3)B}({\bf r}_1, {\bf r}_2, \ldots, {\bf r}_N; 
{\bf r}_1+{\bf s}) \biggl)
\,\exp\biggl(\frac{i}{v^{(3)}} 
\sum_{\beta=3,4} \sum_{j\in\beta \atop j\neq 1}
\int_0^s ds' \Delta V^{(3\beta)}({\bf r}_{1j}, {\bf s}')\biggl)
\Biggl] 
\nonumber \\ [2mm]
& - & e^{is\omega_q^{(3)}/v^{(3)}} \frac{1}{\rho_3} \rho_{1D}(s) \ ,
\label{resp3he-e3}
\end{eqnarray}
as now the term inside the square brackets admits a cumulant
expansion. Up to the first order this leads to
\begin{eqnarray}
\Delta S^{(3)}(q,s) & = & e^{is\omega_q^{(3)}/v^{(3)}} 
\frac{1}{\rho_{1D}(s)} \rho_1^{(3)}(s)
\label{resp3he-f1} \\ [2mm]
& \times & \left[ \exp\left[ -\frac{1}{\rho_{1D}(s)} \int d{\bf r}
\Delta\rho_2^{(3,3)}({\bf r}, 0; {\bf r}+{\bf s}) 
\left[1-\exp\left(\frac{i}{v^{(3)}} \int_0^s ds'
\Delta V^{(33)}({\bf r}, {\bf s})\right)\right] \right. \right.
\nonumber \\ [2mm]
& & \left. -\frac{1}{\rho_{1D}(s)} \int d{\bf r}
\Delta\rho_2^{(3,3)}({\bf r}, 0; {\bf r}+{\bf s}) 
\left[1-\exp\left(\frac{i}{v^{(3)}} \int_0^s ds'
\Delta V^{(33)}({\bf r}, {\bf s})\right)\right] -1 \right] \ .
\nonumber
\end{eqnarray}

The results in Eqs.~(\ref{resp3he-c1}),~(\ref{resp3he-d1}),~(\ref
{resp3he-d2a}),~(\ref{resp3he-d2b}) and~(\ref{resp3he-f1}) constitute
the prediction of the $^3$He response of the mixture in the present
formalism. As before, the FSE functions are built upon the interatomic
potentials and the semidiagonal two--body density matrices, even
though the latter must now be computed for both the real mixture and
its boson--boson approximation. As in the $^4$He case, these results
resemble the original ones derived by Gersch and Rodriguez and form in
fact their extension to the mixture when the problem of finding a
useful prescription for the $^3$He response is addressed.

The $^4$He and $^3$He FSE functions can be evaluated once one has a
suitable description of the required ground state two--body density
matrices. These can be computed starting from a variational Jastrow
wavefunction in the framework of the HNC/FHNC equations for the
mixture, generalizing the formalism developed in
Refs.~\cite{RisCla,RisCla2}.  In the simplest scheme one disregards
the contribution of the Abe diagrams and works in the Average
Correlation Approximation (ACA), in which the $(4,4)$, $(4,3)$ and
$(3,3)$ correlation factors are assumed to be equal. In this limit,
differences between the isotopes are left to the effects derived from
their different mass and statistics. Despite the simplifications, the
wavefunction generated in this way still captives the essential
features of the mixture at $T=0$.

In the particular case of the $^3$He--$^4$He mixture, the interaction
does not distiguish between isotopes and so all three potentials
$V^{(\alpha,\beta)}(r)$ have been taken to be equal to the HFDHE2 Aziz
potential~\cite{Aziz}. The computed FSE functions of the $x_3=0.095$
$^3$He concentration mixture at equilibrium density
$\rho=0.3554\,\sigma^{-3}$ are shown in figure~(1).  The $^4$He
($^3$He) FSE function has been Fourier transformed from $s$ to the
$Y_4$ ($Y_3$) West scaling, where $Y_\alpha=m_\alpha\omega/q-q/2$.
The solid and dashed lines on the left show $R^{(4)}(q,Y_4)$ compared
to the Gersch--Rodriguez FSE function of pure
$^4$He~\cite{MazzBoPo-1,FerPhD} at saturation density
($\rho=0.365\,\sigma^{-3}$). The solid, dashed and dotted lines on the
right depict $R^{(3)}(q,Y_3)$, $\Delta S^{(3)}(q,Y_3)$ and the $^3$He
Compton profile $J^{(3)}(Y_3)=(q/m_3) S^{(3)}(q,Y_3)$ in the IA. As it
can be seen and despite the different densities, $R^{(4)}(q,Y_4)$ is
quite similar to the pure $^4$He FSE function, the main differences
being present at the tails. On the other hand, $R^{(3)}(q,Y_3)$ is
close to both functions, even though the peak is slightly higher. This
effect is mostly due to the low partial density of the $^3$He
component at the mixture $\rho_3=x_3 \rho=0.0338\,\sigma^{-3}$,
compared with the partial density of the $^4$He component
$\rho_4=(1-x_3)\rho=0.3216\,\sigma^{-3}$ and the equilibrium density
of pure $^4$He. Moreover, the $^3$He additive term is rather small
compared with the IA prediction, thus indicating that at high $q$
statistical corrections to the purely bosonic FSE in the $^3$He peak
are present but play a much less significant role compared to the
effect produced by $R^{(3)}(q,Y_3)$.

In summary, we have presented a new formulation of FSE for
$^3$He--$^4$He mixtures, where the presence of a fermionic component
forbids a straightforward application of most of the existing FSE
theories used in the analysis of the high momentum transfer response
of pure $^4$He. The formalism is inspired in the theoretical analysis
carried out by Gersch and Rodriguez, and actually reduces to it in the
zero $^3$He concentration limit. The resulting expressions are
expected to accurately describe the effect of FSE in the response of
the mixture. Furthermore, numerical calculations reveal that the
$^4$He FSE function is similar to the pure $^4$He one. In this way,
FSE corrections to the $^4$He peak in the mixture and in the pure
phase are expected to be similar. Statistical effects do not
appreciably modify a picture in which the FSE of the $^3$He
component are computed in the boson--boson approximation.


\acknowledgments

The author would like to thank Prof.A.S.Rinat for valuable comments on
the $1/v^{(\alpha)}$ expansion of the response. Profs.A.Polls and
J.Boronat are also acknowledged for helpful discussions on the
diagrammatic expansion of the two--body density matrix of the mixture.


\begin{figure}
\caption{Left: $R^{(4)}(q,Y_4)$ compared to the pure $^4$He FSE
function of Gersch and Rodriguiez (solid and dashed lines). Right:
$R^{(3)}(q,Y_3)$ and $\Delta S^{(3)}(q,Y_3)$ compared to the IA
prediction for the $^3$He peak in the mixture (solid, dashed and
dotted lines, respectively).}
\end{figure}



\begin{references}


\bibitem{GerRodSmi} H.A.Gersch, L.J.Rodriguez, amd P.N.Smith, 
	Phys.Rev. {\bf A5}, 1547 (1972).

\bibitem{Rinat-1} A.S.Rinat, Phys.Rev. {\bf B40}, 6625 (1989).

\bibitem{Glyde-1} H.R.Glyde, Phys.Rev. {\bf B50}, 6726 (1994).

\bibitem{MazzBoPo-1} F.Mazzanti, J.Boronat, and A.Polls, 
	Phys.Rev. {\bf B53}, 5661 (1996).

\bibitem{CowWod} R.A.Cowley, and A.D.B.Woods, 
	Phys.Rev.Lett. {\bf 21}, 787 (1968).

\bibitem{MarSweWod} P.Martel, E.C.Swenson, A.D.B.Sears, V.F.Woods, and 
	R.A.Cowley, J.LowTemp.Phys. {\bf 23}, 285 (1976).

\bibitem{SosSnoSok-1} T.R.Sosnick, W.M.Snow, P.E.Sokol, and
	R.N.Silver, Europhys.Lett. {\bf 9}, 707 (1989)

\bibitem{KHAnd} K.H.Anderson. W.G.Stirling, and H.R.Glyde,
        Phys.Rev. {\bf B56}, 8978 (1997).

\bibitem{RTAzuah} R.T.Azuah, W.G.Stirling, H.R.Glyde, M.Boninsegni, P.E.Sokol,
        and S.M.Bennington, Phys.Rev. {\bf B56}, 14620 (1997).

\bibitem{WanSok} Y.Wang and P.E.Sokol, Phys.Rev.Lett. {\bf 72}, 
	1040 (1994).

\bibitem{BorPoFab} J.Boronat, A.Polls, and A.Fabrocini, J.Low
        Temp.Phys. {\bf 91}, 275 (1993).

\bibitem{BonCep} M.Boninsegni and D.M.Ceperley,
        Phys.Rev.Lett. {\bf 74}, 2288 (1995).

\bibitem{Sokmomdist} {\it Momentum Distributions}, edited by R.N. Silver and
	P.E. Soko l (Plenum, New York, 1989).

\bibitem{Glyde-94} H.R.Glyde, in {\em Excitations in Liquid and Solid
        Helium}, chaps.18 and 19, Oxford University Press, New York 
	(1994).

\bibitem{Sears-86} V.F.Sears in ``Neutron Scattering'' Vol 23A of
        ``Methods of Experimental Physics'' K.S\"old and D.L.Price,
        Academic Press, New York (1986) p.521.

\bibitem{GerRod} H.A.Gersch and L.J.Rodriguez, Phys.Rev. {\bf 8}, 905
        (1973). 



\bibitem{RisCla} M.L.Ristig and J.W.Clark, Phys.Rev. {\bf B41}, 8811
        (1990). 

\bibitem{FerPhD} F.Mazzanti, PhD thesis, Universitat de Barcelona
        (1997). 

\bibitem{RisCla2} M.L.Ristig and J.W.Clark, Phys.Rev. {\bf B40}, 4355
        (1989). 

\bibitem{Aziz} R.A.Aziz, V.P.S.Nain, J.S.Carley, W.L.Taylor and
        G.T.McConville. J.Chem.Phys. {\bf 70}, 4330, (1979).

\end{references}
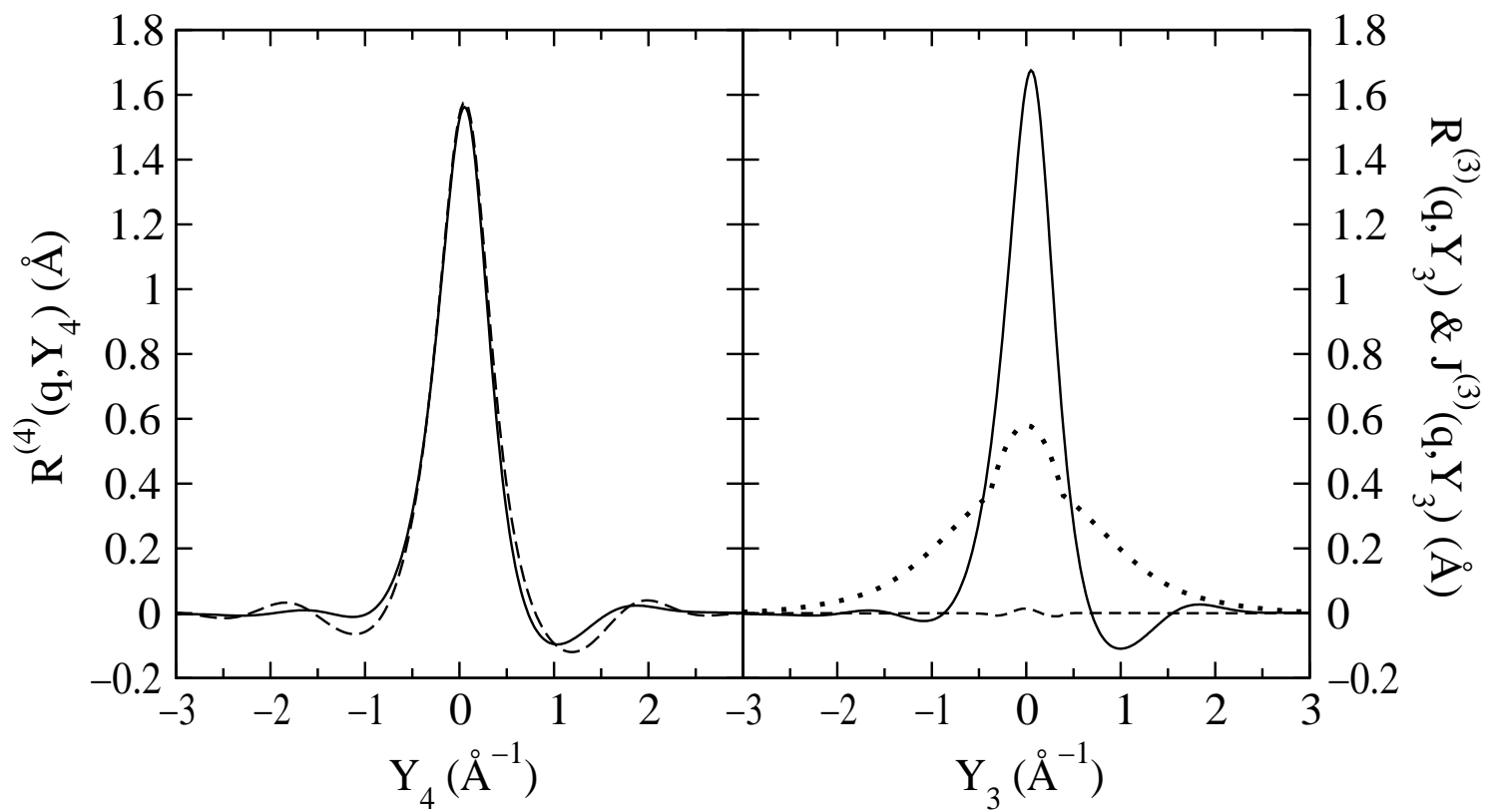
\end{document}